\documentclass[11pt]{article}

\usepackage{amssymb,amsmath,amsfonts}
\usepackage{graphicx}
\usepackage{graphics}
\usepackage{eepic,epsfig}


\begin{document}

\newcommand{\tc}{\textcolor}
\newcommand{\g}{blue}
\newcommand{\ve}{\varepsilon}

\title{The Holographic cosmology with axion field }

\author{A. A. Saharian$^{1}$, A. V. Timoshkin$^{2,3,4}$\\
\\
\textit{$^1$Department of Physics, Yerevan State University,}\\
\textit{Alex Manoogian Street, 1, 0025 Yerevan, Armenia}\\
\textit{$^2$Tomsk State Pedagogical University,}\\
\textit{Kievskaja Street, 60, 634061 Tomsk, Russia}\\
\textit{$^3$International Laboratory of Theoretical Cosmology}\\
\textit{Tomsk State University of Control Systems and Radio Electronics,}\\
\textit{Lenin Avenue, 36, 634050 Tomsk, Russia}\\
\textit{$^4$Kazan (Volga) Federal University, Kremlin Street, 18, 420008 Kazan, Russia}}
\maketitle

\begin{abstract}
In present article we consider an axion F(R) gravity model and described with the help of holographic principle the cosmological models of viscous dark fluid coupled with axion matter in a spatially flat Friedmann-Robertson-Walker (FRW) universe. This description based on generalized infrared-cutoff holographic dark energy, proposed by Nojiri and Odintsov. We explored the Little Rip, the Pseudo Rip, and the power-law  bounce cosmological models  in terms of the parameters of the inhomogeneous equation of the state of viscous dark fluid and calculated the infrared cutoffs analytically. We represented the energy conservation equation for the dark fluid from a holographic point of view and showed a correspondence between the cosmology of a viscous fluid and holographic cosmology. We analyzed the autonomous dynamic system. In the absence of interaction between fluids, solutions are obtained corresponding to two cases. In the first case, dark energy is missing and the extension describes the component of dark matter. The second case corresponds to cosmological models with an extension due to dark energy. The solutions obtained are investigated for stability. For a cosmological model with the interaction of a special type, the stability of solutions of the dynamic system is also investigated.

\end{abstract}


\bigskip

\section{Introduction}

One of the interesting approaches to describing the accelerated expansion of the early and late Universe is the use of the holographic principle [1]. The origin of the holographic principle is associated with the thermodynamics of black holes and string theory [2, 3]. Until recently, studies with the use of holographic principle were carried out mainly in the cosmology of the late-time Universe, in the dark energy era. The generalized cutoff holographic dark energy model was introduced by Nojiri and Odintsov [4, 5].
There are several choices for the infrared cutoff. It can be identified with a combination of FRW universe parameters: the
Hubble function, the particle and the future event horizons, the cosmological constant, and the finite life time of the universe.
A general infrared cutoff may be constructed as an arbitrary combination all above quantities and their derivatives. If the life-time of the universe is finite due to a Big Rip singularity, then the infrared radius depends on the singularity time. Different generalized holographic dark energies corresponding to various versions of the cutoffs were considered in articles \cite{6, 7, 8, 9, 10, 11, 12, 13, 14, 15}. The holographic scenario was recently applied to the inflationary Universe \cite{16,17},  and to  bouncing cosmology \cite{18, 19}.  The theory of  holographic description of the early and late universe is of phenomenological interest because it is  supported  by the  astronomical observations \cite{20, 21, 22, 23}.  Various approaches of the investigation of dark energy have been demonstrated  in the reviews \cite{24, 25}.

In this article we will obtain the holographic picture of the late-time Universe and of the bounce cosmology, associating infrared cutoff with particle and future horizons, and construct the corresponding forms of the energy conservation equations. We will suppose that the main component of cold dark matter in the universe is the axion \cite{26}.  We will study the Little Rip, Pseudo Rip and bounce power-law models,  based on a viscous dark fluid coupled with axion matter. Thereby, the equivalence between these models will be   establish both in the form of a viscous dark fluid coupled with dark matter, and in a holographic form.

\section{The holographic viscous dark fluid coupled with axion matter}

We give the main positions of the holographic principle, following the terminology M. Li \cite{1}.
According to the holographic principle, all physical quantities inside the universe, including the density of dark energy, can be described by some values on the space-time boundary \cite{27}. The density of holographic dark energy can be calculated using a parameters: mass  Plank $M_p$  and some characteristic length $L_{IR}$ (infrared radius) \cite{1}
 \begin{equation}
 \rho=  3c^2M_p^2L_{IR}^{-2},\label{1}
 \end{equation}
 where $c$ is a parameter (dimensionless in geometric units).
According to model of Nojiri and Odintsov \cite{5}, the holographic  energy density is assumed to be inversely proportional to the square of the infrared cutoff $L_{IR}$,
 \begin{equation}
 \rho=  \frac{3c^2}{k^2L_{IR}^2}. \label{2}
 \end{equation}
 When the dark energy is described in this way it means that the horizon cutoff radius is related to the infrared cutoff. At present, there are no precise recommendations for choosing the $ L_{IR}$  parameter, so the size of the particle horizon $L_p$ or the size of the future horizon $L_f$  can be taken as the infrared radius, which are defined respectively \cite{4}
\begin{equation}
L_p=a\int_0^t \frac{dt}{a}, \quad L_f= a\int_t^\infty \frac{dt}{a}.\label{3}
\end{equation}
In the general case, the infrared cutoff $L_{IR}$ could be a combination of $L_p, L_f$ and their derivatives or also the Hubble function, the scale factor, and its derivatives. Using the infrared radius in this form, we can obtain dynamic equations describing the evolution of the late universe on the holographic language.

Let us consider the Universe filled with viscous dark fluid coupled with dark matter in a homogeneous and isotropic spatially flat FRW metric,
\begin{equation}
ds^2= -dt^2+a^2(t)\sum_{i=1}^3 (dx^i)^2, \label{4}
\end{equation}
where $a(t)$ is a scale factor.

One of the opportunities to describe inflation and the era of dark energy is a model of axion F(R) gravity. This model allows the geometric way to combine the era of inflation with the era of dark energy, considering the axion component of dark matter of the Universe. We will consider an F(R) gravity model in the presence of a misalignment axion canonical scalar field with the approximate scalar potential $V(\phi)\simeq \frac{1}{2}m_a^{2}\phi_i^{2}$, where $m_a$  is the axion mass and $\phi_i$ is the axion scalar.

The canonical equation of motion for axion with scalar potential has the form  \cite{26}
\begin{equation}
\ddot{\phi}+3H\dot{\phi}+m_a^{2}\phi=0. \label{5}
\end{equation}
The equation (\ref{5}) is the equation of decaying oscillations, since the second term  is the friction, so the evolution of the universe describes a damped oscillation that approximately begins when $H\sim m_a$  and lasts until the $H\gg m_a$.

We will evaluate the epoch of oscillations with axion. To do this, choose an oscillatory solution of axion in the form
\begin{equation}
\phi(t)=\phi_iA(t)\cos(m_at), \label{6}
\end{equation}
where $\phi_i$ is the initial value of the field of axion after the end of inflation and $A(t)$ is slow-varying function. The property of the monotony of the function $A (t)$  is determined by the conditions
\begin{equation}
\frac{\dot{A}}{m_a}\sim\frac{H}{m_a}\simeq{\epsilon}\ll1, \label{7}
\end{equation}
which is performing for cosmic times for which  $m_a\gg H$.

Let's substitute the solution (\ref{6}) to equation (\ref{5}), and, given the conditions (\ref{7}). The consequence of this is the approximation of the differential equation (\ref{5}) of the view
\begin{equation}
\frac{dA}{A}=-\frac{da}{a},  \label{8}
\end{equation}
analytical solution of which is as follows
\begin{equation}
A\sim a^{-3/2}. \label{9}
\end{equation}
The energy density and pressure of misalignment the axion field are respectively equal
\begin{equation}
\rho_a=\frac{\dot{\phi^2}}{2}+V(\phi), \label{10}
\end{equation}
\begin{equation}
P_a=\frac{\dot{\phi^2}}{2}-V(\phi). \label{11}
\end{equation}
These relations follow the fact, that the misalignment axion field is a canonical scalar field.
Using solution (\ref{6}), one can calculate the term $\frac{\dot{\phi^2}}{2}$. In approximation (\ref{7}) we will obtain
\begin{equation}
\frac{\dot{\phi^2}}{2}\simeq\frac{1}{2}m_a^2\phi_i^2A^2\sin^2(m_at). \label{12}
\end{equation}
Then, the axion potential is equal to
\begin{equation}
V(\phi)=\frac{1}{2}m_a^2\phi_i^2A^2\cos^2(m_at). \label{13}
\end{equation}
Let's substitute expressions (\ref{12}) and (\ref{13}) in the formula for the axion energy density (\ref{10}). This leads to the result
\begin{equation}
\rho_a\simeq\frac{1}{2}m_a^2\phi_i^2A^2. \label{14}
\end{equation}
Then, considering (\ref{9}), the expression for the axion energy density will take \cite{26}
\begin{equation}
\rho_a\simeq \rho^{(0)}_m a^{-3}, \label{15}
\end{equation}
where $\rho^{(0)}_m=\frac{1}{2}m_a^{2}\phi_i^{2}$.

Thus, the axion energy density behaves in this way $\rho_a\sim a^{-3}$ for all cosmic times for which $m_a\gg H$. Hence axion scalar scales as a cold dark matter perfect fluid.

Using (\ref{12}) and (\ref{13}), one can calculate with help (\ref{11}) the pressure of the axion scalar
\begin{equation}
P_a\simeq\frac{1}{2}m_a^2\phi_i^2A^2[\sin^2(m_at)-\cos^2(m_at)]. \label{16}
\end{equation}
The equation of state parameter for the axion scalar $\omega_a={P_a}/\rho_a$   is equal
\begin{equation}
\omega_a=\sin^2(m_at)-\cos^2(m_at). \label{17}
\end{equation}
Calculating the average value of the thermodynamic parameter in the interval equal to the period of trigonometric functions leads to the result $<\omega_a>=0$. This result confirms the fact that  the axion scalar is a cold dark matter particle. Next, we will assume that the main component of cold dark matter in the universe is axion.

Let's consider the system, namely: dark fluid coupled with axion matter. The first Friedmann equation can be written as \cite{28}
 \begin{equation}
 H^2= \frac{1}{3}k^2(\rho+\rho_a), \label{18}
 \end{equation}
The modified continuity and acceleration equations have the form \cite{28}
\begin{equation}
\dot{\rho}+3H(p+\rho)=-Q, \nonumber
\end{equation}
\begin{equation}
\dot{\rho}_a+3H(p_a+\rho_a)= Q, \label{19}
\end{equation}
\begin{equation}
\dot{H}=-\frac{1}{2}k^2(p+\rho +p_a +\rho_a), \nonumber
\end{equation}
 where $H=\dot{a}/a$ is the Hubble parameter and $k^2=8\pi G$ is Einstein's gravitational constant with Newton's gravitational constant $G$;  $p,\rho$ and $p_a,\rho_a$ are the pressure and the energy density of dark energy and axion matter. The function $Q(t)$ describes the interaction between dark energy and axion matter. A dot is the derivative with respect to the cosmic time $t$.

Next, present the continuity equation in holographic form for various cosmological models of the accelerating universe in the axion F(R) gravity.

\section{Holographic description of accelerating universe in axion F(R) gravity}

Let's suppose, that a viscous dark fluid coupled with axion matter, is associated with a holographic energy density. We will consider the late universe and cyclic cosmology with a rebound in the holographic picture, given the following cosmological models.

\subsection{Little Rip model}

In case of Little Rip cosmology an energy density $\rho$ increase with time asymptotically, that is it takes an infinitly long time for forming singularity. The equation of state parameter $\omega < -1$, but $\omega \rightarrow -1$ asymptotically. We have a soft type singularity.

Let's consider the example of a Little Rip model in which the Hubble function has the form \cite{29}
\begin{equation}
H= H_0 e^{\lambda t}, \quad H_0>0, \quad \lambda >0, \label{20}
\end{equation}
where $H_0=H(0)$, $t=0$ denoting present time.

Let us assume, that the axion matter is a dust matter, then $p_a=0$. Now, the gravitational equation for axion matter has more simple view,
\begin{equation}
\dot{\rho}_a+3H\rho_a=Q. \label{21}
\end{equation}
We will consider the coupling of dark energy with axion matter in the view
\begin{equation}
Q=3 \rho_aH(1-C_1\exp(\frac{2}{\lambda}H)), \label{22}
\end{equation}
when the solution of Eq.~(\ref{7}) for the axion energy density has the form (\ref{5}).

Let's the Little Rip universe satisfies an inhomogeneous equation of state for the viscous fluid  \cite{32}
\begin{equation}
p= \omega(\rho,t)\rho -3H\zeta(H,t), \label{23}
\end{equation}
where $\omega(\rho,t)$ is a thermodynamic parameter and $\zeta(H,t)$ is the bulk viscosity, taken in general to depend on both the Hubble function and on the time $t$. From thermodynamic considerations it follows that  $\zeta(H,t)>0$.

Let's consider the simplest case, when the thermodynamic parameter $\omega(\rho,t)=\omega_0$ and the bulk viscosity $\zeta(H,t)= \zeta_0$ are both constants. Then the equation of state (\ref{23}) will take the view
\begin{equation}
p=\omega_0\rho-3\zeta_0H. \label{24}
\end{equation}
Now, we can calculate the scale factor
\begin{equation}
a(t)= a_0\exp \left( \frac{H_0}{\lambda}\exp (\lambda t)\right) \label{25}
\end{equation}
and the future horizon $L_f$
\begin{equation}
L_f = -\frac{1}{\lambda}\exp\left(- \frac{1}{\lambda}H\right)Ei\left( -\frac{1}{\lambda}H\right), \label{26}
\end{equation}
where $Ei(bx), \, b \neq 0$ is the integral exponential function \cite{34}. If  $t \to {+}\infty$ then $L_f \to 0$.

For the Hubble function $H$ in terms of future horizon $L_f$  and its time derivative, one has \cite{5},
\begin{equation}
H= \frac{\dot{L}_f+1}{L_f}, \quad \dot{H}= \frac{\ddot{L}_f}{L_f}- \frac{\dot{L}_f^2}{L_f^2}-\frac{\dot{L}_f}{L_f^2}. \label{27}
\end{equation}
Let's suppose that $\omega_0=-1$ and $C_1=1$. Then,  using (\ref{22}), (\ref{24})  and (\ref{27}), the energy conservation law in the holographic language takes the form
\begin{equation}
(\frac{\lambda}{k^2}-3\xi_0)\left( \frac{\dot{L}_f+1}{L_f}\right) =
\rho^{(0)}_a\exp\left(- \frac{\dot{L}_f+1}{L_f}\right). \label{28}
\end{equation}

Thus, we obtained the Little Rip representation of the holographic principle for a viscous fluid  coupled with axion matter.

\subsection{Pseudo Rip model}

We will investigate example of the Pseudo Rip cosmology in which the Hubble function tends to a constant value (cosmological
constant or asymptotically de Sitter space) in the far future.

Let us suppose that the Hubble function is given as \cite{29}
\begin{equation}
H= H_1-H_0\exp(-\tilde{\lambda}t), \label{29}
\end{equation}
where $H_0, H_1$ and $\tilde{\lambda}$ are positive constants, $H_0>H_1,\, t>0$. For small time values ($t\rightarrow 0$), we obtain $H \rightarrow H_1-H_0$, and when $t\rightarrow +\infty$ the Hubble function also tends to constant value $H \rightarrow H_1$.

In this case we consider the coupling between dark energy and axion matter in the form
\begin{equation}
Q=3 \rho_aH\left(1-a^2 \right), \label{30}
\end{equation}
 then we obtain the solution of the gravitational equation (\ref{19}) in the view (\ref{15}).

Let's study the case, when the thermodynamic parameter is constant, $\omega(\rho,t)=\omega_0$, and assume, that the bulk viscosity is proportional to $H$, $\zeta(H,t)=3\tau H$. Then the equation of state (\ref{23}) wrought as
\begin{equation}
p= \omega_0\rho -9\tau H^2, \label{31}
\end{equation}
where $\tau$ is a positive dimensional constant.

The calculation of the scale factor leads to the result
\begin{equation}
a(t)= a_0\exp\left[ H_0t+\frac{H_1}{\tilde{\lambda}}\exp(-\tilde{\lambda}
t)\right]. \label{32}
\end{equation}

Let's suppose, that $\tilde{\lambda}=H_0$ and calculate the future horizon $L_f$. We obtain
\begin{equation}
L_f= \frac{1}{H_1}\left\{ \exp\left[\frac{H_1}{H_0}\exp(-H_0t)-1\right]\right\}\exp(H_0t). \label{33}
\end{equation}
In the limit  $t\rightarrow {+}\infty$ one obtains $L_f \rightarrow\frac{1}{H_0}$.

Considering  Eq.~(\ref{27}), Eq.~(\ref{30}) and  Eq.~(\ref{31}) we can rewrite the energy conservation law (\ref{19}) in a holographic language
\begin{equation}
2\left( \frac{\ddot{L}_f}{L_f}-\frac{\dot{L}_f^2}{L_f^2}- \frac{\dot{L}_f}{L_f^2}\right) +
9\tau k^2\left( \frac{\dot{L}_f+1}{L_f}\right)^2 = -\rho^{(0)}_a k^2\exp(-H_0t)\exp(1-\frac{\dot{L}_f+1}{H_0L_f}), \label{34}
\end{equation}
assuming that $\omega_0=-1$ and $a_0=1$.

Thus, we applied the holographic principle in the Pseudo Rip model in order to obtain the appropriate energy conservation law.

\subsection{Bounce cosmology}

Let's turn to the consideration of the cosmology model with a rebound \cite{34,35,36}. This model takes the evolution of the universe from the accelerated collapse phase to the accelerated expansion phase through the rebound without the formation of singularity. We obtain the model of the cyclic universe.

We will describe the bounce model on holographic language, choosing as infrared cutoff the particle horizon, and obtain the representation of the energy conservation law in this term. Accounting for the interaction between fluid components in the holographic picture will allow better agreement  between the theory and data of astronomical observations \cite{32}.

Let's investigate a bounce power-law model with the scale factor $a$  in the form \cite{37},
 \begin{equation}
 a(t)= a_0+\beta(t-t_0)^{2n}, \label{35}
 \end{equation}
 where $a_0, \beta$ are positive dimensional constants, $n \in N$, and $t_0$ a fixed bounce time.

 The Hubble function is
 \begin{equation}
 H(t)= \frac{2n\beta (t-t_0)^{2n-1}}{a_0+\beta(t-t_0)^{2n}}. \label{36}
 \end{equation}

We will leave the coupling with axion matter to have the same form (\ref{30}).
If the thermodynamic parameter is equal $\omega(\rho,t)=A_0\rho-1$ and the bulk viscosity is constant $\zeta(H,t)= \zeta_0$, then we get the following expression for the equation of state (\ref{23})
 \begin{equation}
 p=(A_0\rho-1)\rho-3\zeta_0H. \label{37}
 \end{equation}
 We can calculate the particle horizon for the case $n=1$
 \begin{equation}
 L_p= \frac{1}{\sqrt{\beta a_0}}\left[a_0+\beta(t-t_0)^{2}\right]\left[\arctan\sqrt{\frac{\beta}{a_0}}(t-t_0)+\arctan\sqrt{\frac{\beta}{a_0}}t_0\right]. \label{38}
 \end{equation}
In the limit  $t\rightarrow t_0$ (the bounce time) one obtaines $L_p\rightarrow\sqrt\frac{a_0}{\beta}\arctan\sqrt{\frac{\beta}{a_0}}t_0$.
 From the holographic point of view, $H$ can be represented also by the particle horizon $L_p$ \cite{5},
\begin{equation}
H= \frac{\dot{L}_p-1}{L_p}, \quad \dot{H}= \frac{\ddot{L}_p}{L_p}- \frac{\dot{L}_p^2}{L_p^2}+\frac{\dot{L}_p}{L_p^2}. \label{39}
\end{equation}
 Considering  (\ref{30}), (\ref{37}) and (\ref{39}), we can rewrite the energy conservation law as follows
 \begin{eqnarray}
 \frac{2}{k^2}\left( \frac{\ddot{L}_p}{L_p}-\frac{\dot{L}_p^2}{L_p^2}+ \frac{\dot{L}_p}{L_p^2}\right) +
 A_0\left( \frac{\dot{L}_p-1}{L_p}\right)^4\left[\frac{3}{k^2}-\frac{\rho^{(0)}_m}{8\beta^3(t-t_0)^3}\left( \frac{\dot{L}_p-1}{L_p}\right)\right]^2=\nonumber\\
=\left[\frac{\rho^{(0)}_m}{2\beta(t-t_0)}+3\xi_0\right]\left(\frac{\dot{L}_p-1}{L_p}\right) . \label{40}
\end{eqnarray}
The equation (\ref{40}) shows  the holographic description of the  bounce power-law model with viscous fluid.

\section{Analysis of the autonomous dynamic system}

Here we consider a model of the holographic dark energy and
will take (\ref{1}) with the infrared cutoff $L_{IR}=L_{f}$. For this choice one
gets
\begin{equation}
\rho _{d}^{\prime }=-2\rho _{d}\left( 1-\frac{\sqrt{\Omega _{d}}}{c}\right) ,
\label{rodp}
\end{equation}%
where the prime stands for the derivative with respect to $\ln a$: $%
f^{\prime }=df/d\ln a$ and%
\begin{equation}
\Omega _{d}=\frac{\rho _{d}}{\rho _{c}}=\frac{c^{2}}{L_{IR}^{2}H^{2}},\;\rho _{c}=%
\frac{3H^{2}}{k^{2}}.  \label{Omdn}
\end{equation}%
The derivative with respect to the time coordinate $t$ will be denoted, as
usual, by dot: $\dot{f}=df/dt$. From the Friedmann equation
\begin{equation}
H^{2}=\frac{k^{2}}{3}(\rho _{d}+\rho _{i}),  \label{H2F}
\end{equation}%
with $H=\dot{a}/a$ and $\rho _{i}$ being the energy density for other forms
of the matter, we have $\Omega _{d}+\Omega _{i}=1$, where $\Omega _{i}=\rho
_{i}/\rho _{c}$. The energy density $\rho _{i}$ will include, in particular,
the contributions coming from the dust matter, radiation or cosmological
constant.

Taking the derivative for $\Omega _{d}$, from (\ref{Omdn}) one obtains%
\begin{equation}
\frac{H^{\prime }}{H}=\frac{\sqrt{\Omega _{d}}}{c}-1-\frac{\Omega
_{d}^{\prime }}{2\Omega _{d}}.  \label{Hder1n}
\end{equation}%
From the other side, taking the derivative of the equation (\ref{H2F}) we get%
\begin{equation}
\frac{H^{\prime }}{H}=-\Omega _{d}\left( 1-\frac{1}{c}\sqrt{\Omega _{d}}%
\right) +\frac{\rho _{i}^{\prime }}{2\rho _{c}}.  \label{Hder2n}
\end{equation}%
In order to exclude the derivative $\rho _{i}^{\prime }$ we use the energy
conservation equation.
\begin{equation}
\dot{\rho}_{i}+3H(1+w_{i})\rho _{i}=Q,  \label{EnCons}
\end{equation}%
where $w_{i}=p_{i}/\rho _{i}$, with $p_{i}$ being the matter pressure, and $Q
$ describes the energy transfer between the dark energy and matter. From
this equation it follows that
\begin{equation}
\frac{\rho _{i}^{\prime }}{\rho _{c}}=-3(1+w_{i})\left( 1-\Omega _{d}\right)
+\frac{Q}{H\rho _{c}}.  \label{EnCons2}
\end{equation}%
Substituting this in (\ref{Hder2n}) one finds%
\begin{equation}
\frac{H^{\prime }}{H}=\frac{1}{c}\Omega _{d}^{3/2}+\frac{1+3w_{i}}{2}\Omega
_{d}-\frac{3}{2}(1+w_{i})+\frac{Q}{2H\rho _{c}}.  \label{Hder3n}
\end{equation}%
Combining this with (\ref{Hder1n}) we obtain%
\begin{equation}
\Omega _{d}^{\prime }=2\Omega _{d}\left( 1-\Omega _{d}\right) \left( \frac{%
\sqrt{\Omega _{d}}}{c}+\frac{1+3w_{i}}{2}\right) -\frac{Q}{H\rho _{c}}.
\label{Omder}
\end{equation}

By taking into account that $\dot{H}=HH^{\prime }$, from the equation
\begin{equation}
\dot{H}=-\frac{1}{2}k^{2}(\rho _{d}+p_{d}+\rho _{i}+p_{i}),  \label{Hdot}
\end{equation}%
we find%
\begin{equation}
\frac{H^{\prime }}{H}=-\frac{3}{2}\left[ \frac{p_{d}}{\rho _{c}}%
+1+w_{i}\left( 1-\Omega _{d}\right) \right] .  \label{Hder4}
\end{equation}%
Combining this with (\ref{Hder3n}) one gets
\begin{equation}
p_{d}=-\frac{1}{3}\left( 1+\frac{2}{c}\sqrt{\Omega _{d}}\right) \rho _{d}-%
\frac{Q}{3H}.  \label{pd}
\end{equation}%
This relation describes the equation of state for the dark energy.
Note that for $Q=0$ it does not contain the characteristics of the matter content.

Introducing the notations $x=\sqrt{\Omega _{d}}$ and $y=\ln \left(
H/H_{0}\right) $, the equations (\ref{Hder3n}) and (\ref{Omder}) are written
in the form of the dynamical system%
\begin{eqnarray}
x^{\prime } &=&x\left( 1-x^{2}\right) \left( \frac{x}{c}+c_{i}\right) -\frac{%
Q}{2xH\rho _{c}},  \notag \\
y^{\prime } &=&x^{2}\left( \frac{x}{c}+c_{i}\right) -1-c_{i}+\frac{Q}{2H\rho
_{c}},  \label{DS}
\end{eqnarray}%
where we have introduced the notation%
\begin{equation}
c_{i}=\frac{1+3w_{i}}{2}.  \label{ci}
\end{equation}%

Note that the equations (\ref{DS}) are valid for general case when $w_{i}$
is not constant. The models with nonnegative energy density $\rho _{i}$ for
the matter are described by the phase trajectories in the strip $0\leq x\leq
1$ of the phase plane $(x,y)$. For the further discussion of solutions of
the system (\ref{DS}) we need to specify the function $Q$.

First we consider the simplest case with $Q=0$. The corresponding autonomous
dynamical system, obtained from (\ref{DS}), has special solution $x=0$ and $%
x=1$. For the models with $x=0$ the dark energy is absent and they describe
the expansion with the matter having the energy density $\rho _{i}$. For
barotropic sources one has $w_{i}=\mathrm{const}$ and from the second
equation of (\ref{DS}) we get the standard relation $%
H=H_{0}/(a/a_{0})^{3(1+w_{i})/2}$. The solutions with $x=1$ correspond to
cosmological models with the dark energy as the only source driving the
expansion. From the second equation in (\ref{DS}) it follows that $%
H=H_{0}(a/a_{0})^{1/c-1}$. For $c=1$ this corresponds to the de Sitter
solution $a=a_{0}e^{H_{0}(t-t_{0})}$ and one has $L=1/H_{0}$. For $c\neq 1$
the time dependence of the scale factor is given by
\begin{equation}
a/a_{0}=\left[ H_{0}\left( 1-1/c\right) \left( t-t_{1}\right) \right]
^{1/(1-1/c)}.  \label{atx1}
\end{equation}%
This describes accelerated expansion with $t_{1}<t<\infty $ for $c>1$ and
Big Rip singularity at $t=t_{1}$ for $0<c<1$. For $w_{i}=\mathrm{const}$ and
under the condition $-2/c<1+3w_{i}<0$, the system (\ref{DS}), with $Q=0$,
has another special solution $x=-c_{i}c$. The corresponding solution for the
scale factor coincides with that for the case $x=0$. In the case of sources
with $w_{i}=\mathrm{const}$ for the general solution of the first equation
in (\ref{DS}) with $Q=0$ one gets%
\begin{equation}
a=a_{0}x^{\frac{1}{c_{i}}}(1+x)^{\frac{1}{2(1/c-c_{i})}}(1-x)^{-\frac{1}{%
2(1/c+c_{i})}}|x+c_{a}|^{\frac{1}{c_{i}(c^{2}c_{i}^{2}-1)}}.  \label{ax}
\end{equation}%
For the Hubble function we find%
\begin{equation}
H=\mathrm{const}\cdot x^{-1-\frac{1}{c_{i}}}(1+x)^{\frac{c+1}{2(c_{i}c-1)}%
}(1-x)^{\frac{c-1}{2(c_{i}c+1)}}|x+c_{i}c|^{\frac{1+1/c_{i}}{1-c^{2}c_{i}^{2}%
}}.  \label{Hub1}
\end{equation}%
The time coordinate is expressed in terms of $x$ as
\begin{equation}
t=c\int \frac{dx}{x\left( 1-x^{2}\right) \left( x-x_{i}\right) H}.
\label{te}
\end{equation}%

The formulas (\ref{ax})-(\ref{Hub1}) determine the the functions $H=H(t)$ and $%
a=a(t)$ in parametric form. The special solution $x=0$ is unstable for $%
c_{i}>0$ and stable for $c_{i}<0$. The solution $x=1$ is stable for $%
c_{1}+1/c>0$ and unstable for $c_{1}+1/c<0$. The special solution $x=-c_{i}c$
is always unstable. If this solution is present, the solution $x=0$ is the
future attractor for the general solution in the region $0<x<-c_{i}c$ and
the solution $x=1$ is the future attractor in the region $-c_{i}c<x<1$.

\begin{figure}[tbph]
\begin{center}
\epsfig{figure=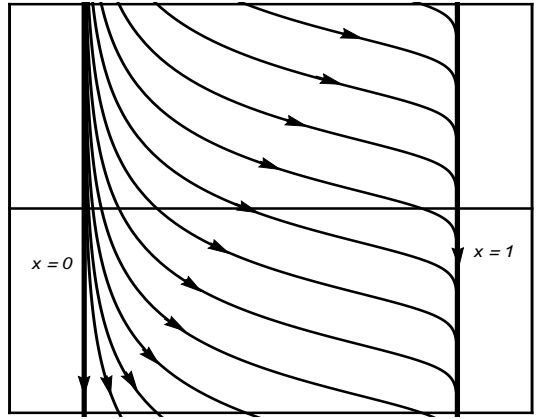,width=7.cm,height=5.5cm}
\end{center}
\caption{The phase portrait of the dynamical system (\ref{DS}) in the absence of the energy transfer between different
components of the energy content for $w_{i}=0$. The phase trajectories are plotted for $c=2.2$.}
\label{fig1}
\end{figure}

For the most important special cases of radiation ($w_{i}=1/3$), dust matter
($w_{i}=0$) and cosmological constant ($w_{i}=-1$) one has $c_{i}=1,1/2,-1$,
respectively. In figure \ref{fig1} we present the phase portrait of the
dynamical system (\ref{DS}) with $Q=0$ in the physical region $0\leq x\leq 1$
of the phase plane $(x,y)$. The phase trajectories are plotted for the dust
matter and for the constant in the expression of the dark energy we have
taken $c=2.2$. As it has been noted above, the special solution $x=0$ is
unstable and the special solution $x=1$ is the future attractor for the
general solution.

\begin{figure}[tbph]
\begin{center}
\begin{tabular}{cc}
\epsfig{figure=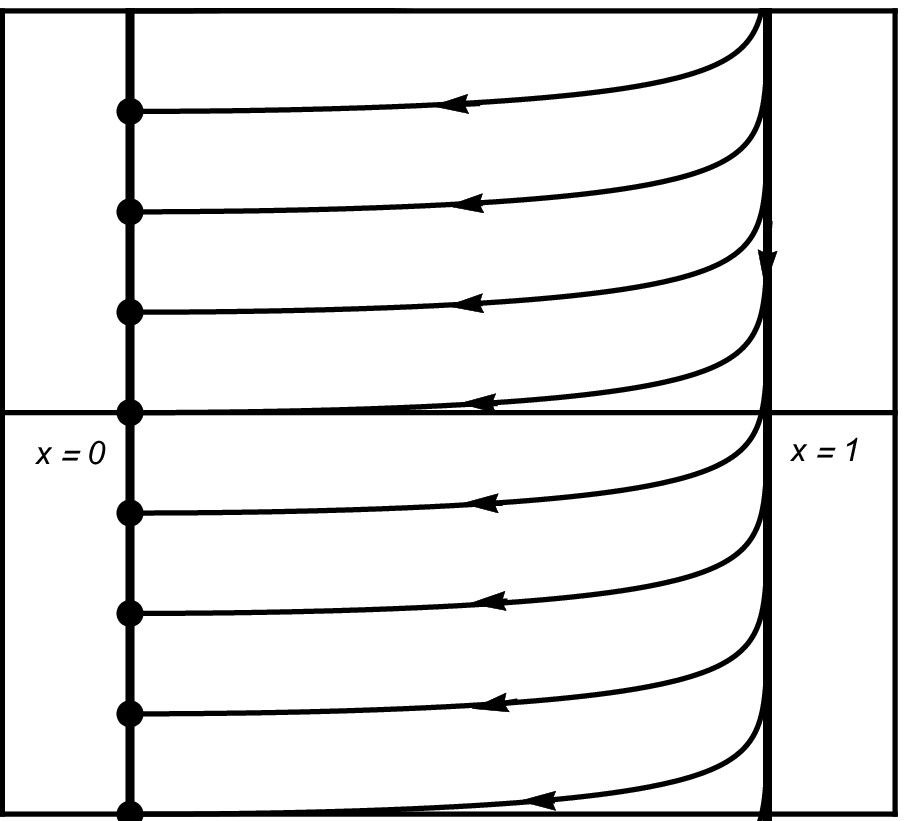,width=7.cm,height=5.5cm} & \quad %
\epsfig{figure=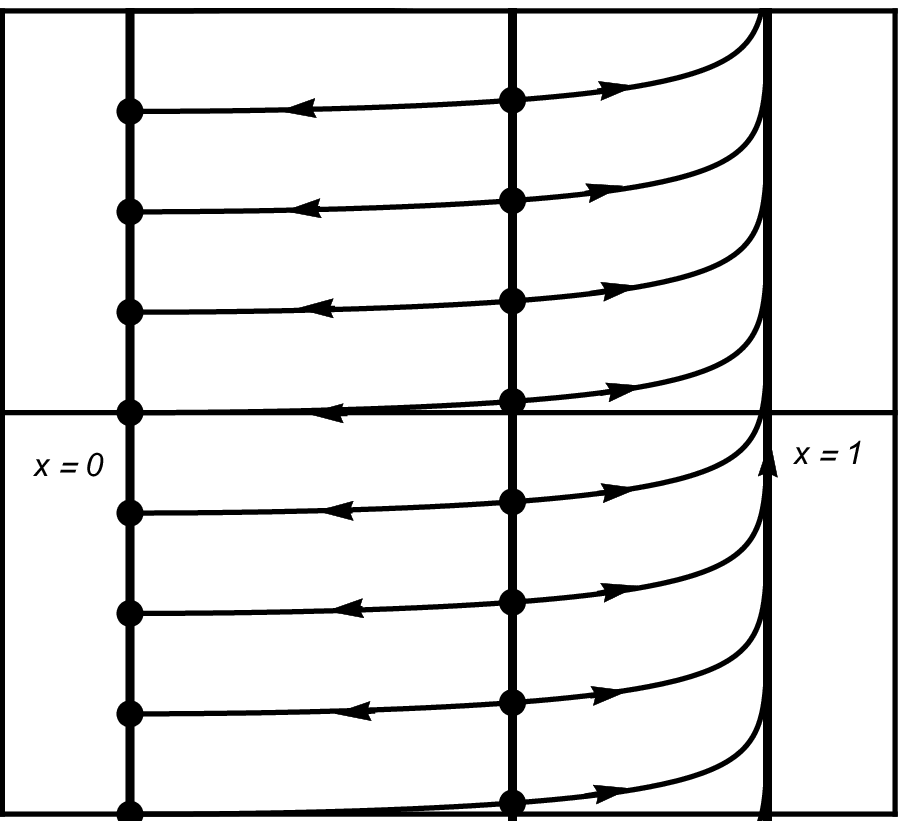,width=7.cm,height=5.5cm}%
\end{tabular}%
\end{center}
\caption{The phase portrait of the dynamical system (\ref{DS}) for $w_{i}=-1$. The left and right
panels correspond to $c=3$ and $c=0.6$, respectively.}
\label{fig2}
\end{figure}

In the case of the cosmological constant the formulas given above for the
model with $Q=0$ are simplified to%
\begin{eqnarray}
H &=&\frac{\mathrm{const}}{\sqrt{1-x^{2}}},  \notag \\
a &=&a_{0}x^{-1}(1+x)^{\frac{1}{2\left( 1/c+1\right) }}(1-x)^{-\frac{1}{%
2\left( 1/c-1\right) }}|x-c|^{\frac{1}{1-c^{2}}}.  \label{HaxCC}
\end{eqnarray}%
In the region $c<1$ there is also a special solution $x=c$ with $H=\mathrm{%
const}$. This corresponds to the de Sitter expansion. In figure \ref{fig2}
we display the phase portraits of the corresponding dynamical system (\ref%
{DS}) for $c=3$ (left panel) and $c=0.6$ (right panel). Note that all the
points on the lines $x=0$ and $x=c$ (in the case $c<1$) are critical points.
The points on the line $x=0$ describe de Sitter expansion driven by the
cosmological constant. For $c>1$ these solutions are future attractors for
the general solution. In the case $c<1$ they serve as attractors for models
with $0<x<c$. The models described by the phase trajectories in the region $%
c<x<1$ tend to the special solution $x=1$ at late stages of the expansion.

As an example with nonzero energy transfer between the dark energy and
matter we consider the model with%
\begin{equation}
Q=\frac{2k^{2}b}{3H}\rho _{i}\rho _{d},  \label{Q1}
\end{equation}
where the factor $2k^{2}/3$ is separated for convenience of the further
discussion. The corresponding dynamical system reads%
\begin{eqnarray}
x^{\prime } &=&\frac{1}{c}x\left( 1-x^{2}\right) \left( x-x_{i}\right) ,
\notag \\
y^{\prime } &=&x^{2}\left[ \frac{x}{c}+c_{i}+b\left( 1-x^{2}\right) \right]
-1-c_{i}.  \label{DS2}
\end{eqnarray}%
with $x_{i}=c(b-c_{i})$. The special solutions corresponding to $x=0$ and $%
x=1$ are the same as in the previous case. For barotropic sources $\rho _{i}$
and under the condition $c_{i}<b<1/c+c_{i}$, the dynamical system (\ref{DS2}%
) has special solution $x=x_{i}$. For this solution from (\ref{DS2}) we get
\begin{equation}
a=a_{0}\left[ H_{0}\beta _{i}\left( t-t_{2}\right) \right] ^{1/\beta
_{i}},\;\beta _{i}=1+c_{i}-bx_{i}^{2}\left( 2-x_{i}^{2}\right) .
\label{atxi}
\end{equation}%
Note that for the transfer function (\ref{Q1}) and for $w_{i}=\mathrm{const}$
the general solution of the first equation in (\ref{DS2}) is presented as%
\begin{equation}
a=a_{0}x^{-\frac{c}{x_{i}}}(1+x)^{\frac{c}{2(x_{i}+1)}}(1-x)^{\frac{c}{%
2(x_{i}-1)}}|x-x_{i}|^{\frac{c}{x_{i}(1-x_{i}^{2})}}.  \label{ax2}
\end{equation}%
From the second equation we get the expression for the Hubble function%
\begin{equation}
H=\mathrm{const}\cdot e^{bcx}x^{\frac{1+c_{i}}{b-c_{i}}}(1+x)^{-\frac{1+c}{%
2(1+x_{i})}}(1-x)^{-\frac{c-1}{2(x_{i}-1)}}|x-x_{i}|^{\gamma _{i}},
\label{Hub}
\end{equation}%
with the notation%
\begin{equation}
\gamma _{i}=\frac{1}{b-c_{i}}\left[ b\left( x_{i}^{2}-1\right) -\frac{%
b-c_{i}-1}{x_{i}^{2}-1}\right] .  \label{gami}
\end{equation}%
For the time coordinate we have the relation (\ref{te}). In combination with
(\ref{ax2}) and (\ref{Hub}), it determines the functions $a=a(t)$ and $H=H(t)
$ in the model under consideration. Note that if the special solution $%
x=x_{i}$ is present ($0<x_{1}<1$) it is always unstable. The solution $x=0$ ($%
x=1$) is the future attractor for the general solution (\ref{ax2}) in the
range $0<x<x_{i}$ ($x_{i}<x<1$). For $x_{i}<0$ or $x_{i}>0$ the special
solution is outside of the physical region $0\leq x_{1}\leq 1$ of the phase
space. In this case the special solution $x=0$ is stable for $x_{i}>1$ and
unstable for $x_{i}<0$. Under the same conditions, the solution $x=1$ is
stable for $x_{i}<0$ and unstable for $x_{i}>1$.

As the main component of cold dark matter we can take the axion field. Under
the condition $H\ll m_{a}$, with $m_{a}$ being the axion mass, the axion
energy density scales as \cite{26} $\rho _{i}=\rho _{a}\sim a^{-3}$.

\section{Conclusion}

In this paper we have considered examples of Little Rip, Pseudo Rip, and bounce cosmology from the standpoint of the holographic principle. In accordance with spatially flat  Friedmann-Robertson-Walker metric, assuming that the universe is filled with viscous dark fluid coupled with axion matter, we reformulated for listed models energy conservation equation on holographic language.
 Thus, we showed the equivalence between viscous fluid cosmology and holographic fluid cosmology with the specific cutoff introduced by Nojiri and Odintsov \cite{4,5}. The stability of solutions obtained for the autonomous dynamic system for cases both with the interaction between the components of two fluids and without interaction between them are investigated. It is shown that in case without interaction, the accelerated expansion of the universe may be due to only dark energy or dark matter. In the second case, the solution received is unstable. For a cosmological model with interaction between the components of the fluids of a special type, solutions of a dynamic system, which have both the property of instability and solutions in the form of an attractor are obtained.

The question arises on the confirmation of the holographic theory of astronomical observations. Such a comparative analysis was carried out in Ref.~\cite{38}. As an example was considered the theoretical holographic dark energy model on the brane. Comparative analysis conducted for the relationships between apparent magnitudes and redshifts for distant supernova Ia, Hubble parameters for different redshifts, and baryon acoustic oscillations confirmed a qood agreement between observed data and theoretical predictions.

\section*{Acknowledgement}

A.A.S. was supported by the grant No. 20RF-059 of the Committee of Science of the Ministry of Education, Science, Culture and Sport RA.  A.V.T. was supported  by Russian Foundation for Basic Research, Project No. 20-52-05009.

\end{document}